\documentstyle[psfig,times]{jfm}

\begin{document}
\newcommand{\1}{\begin{equation}}
\newcommand{\2}{\end{equation}}
\newcommand{\3}{\begin{eqnarray}}
\newcommand{\4}{\end{eqnarray}}
\title
[An Improved Exact Riemann Solver for Relativistic Hydrodynamics]
{An Improved Exact Riemann Solver for Relativistic Hydrodynamics}

\author
[L. Rezzolla \& O. Zanotti]
{L\ls U\ls C\ls I\ls A\ls N\ls O\ls \ns R\ls E\ls Z\ls Z\ls
O\ls L\ls L\ls A$^{1,2}$\ns \and O\ls L\ls I\ls N\ls D\ls
O\ls \ns Z\ls A\ls N\ls O\ls T\ls T\ls I$^1$}

\affiliation{$^1$ SISSA, International School for Advanced Studies,
		Trieste, Italy\\[\affilskip]
	$^2$ INFN, Department of Physics, Trieste, Italy} 

\date{\today}

\maketitle

\begin{abstract}
A Riemann problem with prescribed initial conditions will
produce one of three possible wave patterns corresponding
to the propagation of the different discontinuities that
will be produced once the system is allowed to relax. In
general, when solving the Riemann problem numerically,
the determination of the specific wave pattern produced
is obtained through some initial guess which can be
successively discarded or improved. We here discuss a new
procedure, suitable for implementation in an exact
Riemann solver in one dimension, which removes the
initial ambiguity in the wave pattern. In particular we
focus our attention on the relativistic velocity jump
between the two initial states and use this to determine,
through some analytic conditions, the wave pattern
produced by the decay of the initial discontinuity. The
exact Riemann problem is then solved by means of
calculating the root of a nonlinear
equation. Interestingly, in the case of two rarefaction
waves, this root can even be found analytically. Our
procedure is straightforward to implement numerically and
improves the efficiency of numerical codes based on exact
Riemann solvers.
\end{abstract}

\section{Introduction}
\label{intro}

	The general Riemann problem is based on the
calculation of the one-dimensional temporal evolution of
a fluid which, at some given initial time, has two
adjacent states characterized by different values of
uniform velocity, pressure and density. From a
mathematical point of view, the Riemann problem is an
initial value problem for a hyperbolic system of partial
differential equations, with initial conditions
characterized by a discontinuity between the two initial
states. These initial conditions establish the way in which the
discontinuity will decay after removal of the barrier
separating the initial ``left'' and ``right'' states. The
schematic evolution of a general Riemann problem can be
represented as (Mart\'{\i} \& M\"uller 1994)
\begin{equation}
\label{general}
L{\cal W}_\leftarrow L_*{\cal C}R_*{\cal W}_\rightarrow R
	\ ,
\end{equation}
where ${\cal W}$ denotes a shock or a rarefaction wave
that propagates towards the left $(\leftarrow)$ or the
right $(\rightarrow)$ with respect to the initial
discontinuity, $L$ and $R$ are the known initial left and
right states, while $L_*$ and $R_*$ are the new
hydrodynamic states that form behind the two waves
propagating in opposite directions. These waves are
separated by a contact discontinuity ${\cal C}$ and
therefore have the same values of the pressure and
velocity, but different values of the
density\footnote{Note that the contact discontinuity
might be also trivial so that the density may not be
discontinuous.}. The Riemann problem is said to be
solved when the velocity, pressure and density in the new
states $L_*$ and $R_*$ have been computed, as well as the
positions of the waves separating the four states. The
solution of the one dimensional Riemann problem in
relativistic hydrodynamics was discussed in the general
case by Mart\'{\i} \& M\"uller (1994) and the reader is
referred to their work for further details (see also Pons
{\it et al.} 2000, for the extension to multidimensions).

	The numerical solution of a Riemann problem is
the fundamental building block of hydrodynamical codes
based on Godunov-type finite difference methods. In such
methods, the computational domain is discretized and each
interface between two adjacent grid-zones is used to
construct the initial left and right states of a ``local
Riemann problem''. The evolution of the hydrodynamical
equations is then obtained through the solution across
the computational grid of the sequence of local Riemann
problems set up at the interfaces between successive
grid-zones (see Mart\'{\i} \& M\"uller 1996, but also 
Godunov 1959 and Colella \& Woodward 1984).  
The ``core'' of each of these local Riemann problems consists
of determining the fluid pressure across the contact
discontinuity, which can be calculated by imposing the
continuity of the normal component of the fluid velocity
across ${\cal C}$
\begin{equation}
\label{vls=vrs}
v_{L_*}(p_*)=v_{R_*}(p_*)\ .
\end{equation}
In general, (\ref{vls=vrs}) is a nonlinear algebraic
equation in the unknown pressure $p_*$ and requires a
numerical solution. Depending on the different wave
patterns forming after the decay of the
discontinuity, a different nonlinear equation will need
to be solved\footnote{This approach is usually referred
to as an ``exact'' Riemann solver to distinguish it from
the family of so called ``approximate'' Riemann solvers,
where the system of equations to be solved is reduced to
quasi-linear form, thus avoiding any iterative procedure.  See, for
example, the approximate Riemann solver of Roe (1981).}.
This initial ``ambiguity'' in the wave pattern produced
corresponds to the fact that the interval in
pressure bracketing the solution $p_*$ is not known a priori. 
In practice this lack of information
is compensated by the use of efficient
numerical algorithms which, via a process of trial and error, determine
the correct wave pattern and then proceed to the solution
of the corresponding nonlinear equation (Mart\'{\i}, \&
M\"uller 1999).

	In this paper, we show that the relativistic
expression for the relative velocity between the two
initial states is a function of the unknown pressure
$p_*$ and so a new procedure for numerically solving the
exact Riemann problem can be proposed in which the
pressure $p_*$ is no longer obtained by the solution of
equation (\ref{vls=vrs}). Rather, $p_*$ is calculated by
equating the relativistic invariant expression for the
relative velocity between the two initial states with the
value given by the initial conditions. 

	When compared to equivalent approaches, our exact
Riemann solver has some advantages. Firstly, we can
remove the ambiguity mentioned above and determine the
generated flow pattern by simply comparing the relative
velocity between the two initial states with reference
values built from the initial conditions of the Riemann
problem. Doing so provides immediate information about
which of the nonlinear equations (one for every wave
pattern) needs to be solved.  Secondly, by knowing the
wave pattern we can produce an immediate bracketing of
the solution. Doing so gives improved efficiency in the
numerical root finding procedure. Finally, for one of the
wave patterns (i.e. for two rarefaction waves moving in
opposite directions) our method provides the solution of
the relativistic Riemann problem in a closed analytic
form. 

	Because of its simplicity, the numerical
implementation of our method is straightforward and can
be accomplished with a much smaller number of lines of
code. When compared with other exact Riemann solvers
(e. g. Mart\'{\i} \& M\"uller 1999) it has also proved to
be computationally more efficient. In particular, when
solving a generic hydrodynamical problem (in which one
solves for very simple Riemann problems) the approach
proposed here brackets the solution very closely and this
produces substantial computational improvements of up to
$30\%$. For the cases discussed here however, where very
strong shocks are considered, the speed-up is below
$10\%$.

        This paper briefly introduces our idea and is
organized as follows: in Section~\ref{setup} we define
the mathematical set-up for the formulation of the
Riemann problem in relativistic hydrodynamics. In
Section~\ref{lrvs} we determine the basic relativistic
expressions linking the velocities ahead of and behind a
shock or a rarefaction wave. These expressions will be
used repeatedly in the following
Sections~\ref{case_i}--\ref{case_iii}, where will write
explicit criteria for the occurrence of the three
possible wave patterns. Section~\ref{ni} discusses how
the criteria can be used for making an efficient
numerical implementation of a Riemann solver and briefly
presents a comparison of performances with more
traditional algorithms. Finally, Section~\ref{conclusion}
presents our conclusions. Throughout, we use a system of
units in which $c = 1$.

\section{Setting up the Relativistic Riemann Problem}
\label{setup}

	Consider a perfect fluid with four-velocity
$u^\mu = W(1,v,0,0)$ with $W\equiv(1-v^2)^{-1/2}$ being
the Lorentz factor and with a stress-energy tensor
\begin{equation}
T^{\mu\nu}=(e+p)u^\mu u^\nu+p \eta^{\mu\nu}
	= \rho h u^\mu u^\nu+p \eta^{\mu\nu} \ ,
\end{equation}
where $\mu=0,\ldots,3$, $\eta^{\mu\nu}={\rm
diag}(-1,1,1,1)$ and $e$, $p$ are the proper energy
density and pressure, respectively. Assume moreover that
the fluid obeys a polytropic equation of
state  
\begin{equation}
\label{pol_eos}
p = k(s)\rho^{\gamma}=(\gamma-1)\rho\epsilon \ ,
\end{equation}
where $\rho$ is the proper rest mass density, $\gamma$ is
the adiabatic index, and $k(s)$ is the polytropic
constant, dependent only on the specific entropy $s$ (the
latter is generally assumed to be different in the two
initial states). Straightforward expressions can be
written relating $e$ and $p$ to the specific enthalpy $h$
and to the specific internal energy $\epsilon$ of the
fluid
\begin{eqnarray}
e&=&\rho(1+\epsilon)=\rho+\frac{p}{\gamma-1} \ ,
\\
h&=&1+\epsilon+\frac{p}{\rho}=1+
	\frac{p}{\rho}\left(\frac{\gamma}{\gamma-1}\right)
	\ .
\end{eqnarray}
Consider now the fluid to consist of an initial left
state (indicated with an index $1$) and an initial right
state (indicated with an index $2$), each having
prescribed and different values of uniform pressure,
density and velocity\footnote{Note that hereafter we will
consider the fluid to be of the same type in the two
initial states. In principle, in fact, the fluid in the
two initial states might be governed by two different
equations of state or by polytropic equations of state
with different polytropic indices. The formulation of the
problem in that case is equivalent to the one presented
here for a single type of fluid but particular attention
must be paid to distinguishing the different components
in the relevant expressions.}. The discontinuity between
the two states which has been constructed in this way
will then decay producing one of the three wave patterns
listed below

	{\it (i)~} {\sl two shock waves}, one moving
	towards the initial left state, and the other
	towards the initial right state: $L{\cal
	S}_\leftarrow L_*{\cal C}R_*{\cal S}_\rightarrow
	R$.

	{\it (ii)~} {\sl one shock wave and one rarefaction
	wave}, the shock moving towards the initial right
	state, and the rarefaction towards the initial left
	state (or viceversa if $p_2>p_1$): $L{\cal
	R}_\leftarrow L_*{\cal C}R_*{\cal S}_\rightarrow
	R$.

	{\it (iii)~} {\sl two rarefaction waves}, one
	moving towards the initial left state, and the
	other towards the initial right state: $L{\cal
	R}_\leftarrow L_*{\cal C}R_*{\cal R}_\rightarrow
	R$. A special case of this wave pattern is
	produced when the two rarefaction waves leave a
	vacuum region behind them.
 
	The basis of our approach lies in the possibility
of determining ``a priori'' which of these three wave
patterns will actually result by simply comparing the
value of the special relativistic relative velocity
between the initial left and right states
\begin{equation}
\label{vrel}
v_{12} \equiv \frac{v_1-v_2}{1-v_1 v_2} \ ,
\end{equation}
with the values of the limiting relative velocities for
the occurrence of the wave patterns {\it (i)--(iii)}. In
this respect, our approach represents the relativistic
generalization of a similar analysis proposed in
Newtonian hydrodynamics by Landau and Lifshitz (1987)
(see also Gheller 1997 for a numerical implementation).
Mathematically, the occurrence of the different cases can
be distinguished because the relative
velocity\footnote{For compactness we will hereafter refer
to as ``relative velocity'' the relativistic invariant
expression.} (\ref{vrel}) is a continuous monotonic
function of the pressure $p_*$ across the contact
discontinuity. This is shown in Fig.~\ref{fig1} where we
have plotted $v_{12}$ as a function of $p_*$ and it is
proved mathematically in Appendix A. Note that the curve
shown in Fig.~\ref{fig1} is effectively produced by the
continuous joining of three different curves describing
the relativistic relative velocity for the three
different wave patterns corresponding respectively to two
shocks (2S, indicated as a dashed line), one shock and
one rarefaction wave (SR, indicated as a dotted line),
and two rarefaction waves (2R, indicated as a continuous
line). The joining of the different curves occurs at
precise values of the relative velocity which we have
indicated as $({\tilde v}_{12})_{_{2S}}$, $({\tilde
v}_{12})_{_{SR}}$, $({\tilde v}_{12})_{_{2R}}$ and which
depend uniquely on the initial conditions of the two
unperturbed states. These three values of the relative
velocity mark the extremes of three different intervals
on the vertical axis and correspond to the three
different cases {\it (i)--(iii)}.

\begin{figure}[htb]
\centerline{
\psfig{file=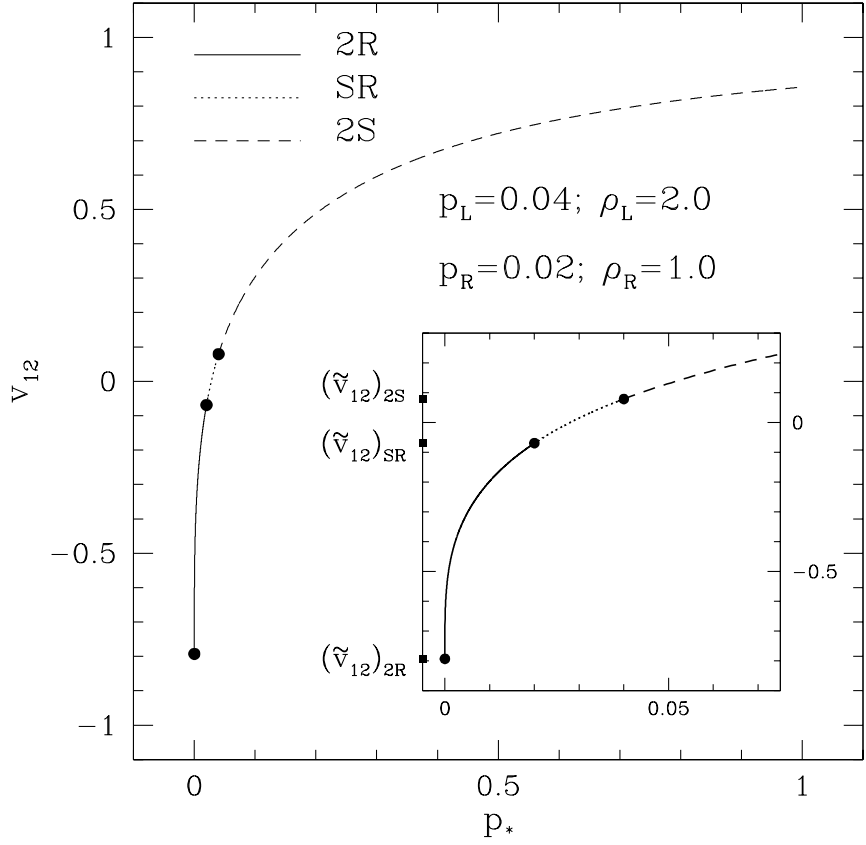,angle=0,width=12.0cm}
        }
\caption{
\label{fig1}
Relative velocity between the two initial states 1 and 2
as a function of the pressure at the contact
discontinuity. Note that the curve shown is given by the
continuous joining of three different curves describing
the relative velocity corresponding respectively to two
shocks (a dashed line), one shock and one rarefaction
wave (dotted line), and two rarefaction waves (continuous
line). The joining of the curves is indicated with filled
dots. The small inset on the right shows a magnification
for a smaller range of $p_*$ and we have indicated with
filled squares the limiting values for the relative
velocities $({\tilde v}_{12})_{_{2S}}$, $({\tilde
v}_{12})_{_{SR}}$, $({\tilde v}_{12})_{_{2R}}$}
\end{figure}

	Within this framework then, it is sufficient to
compare the relative velocity between the initial states
(\ref{vrel}) with the calculated values of the limiting
relative velocities $({\tilde v}_{12})_{_{2S}}$,
$({\tilde v}_{12})_{_{SR}}$, $({\tilde v}_{12})_{_{2R}}$,
to determine, in advance, which wave pattern will be
produced. This, in turn, determines the correct equation
to solve and the correct bracketing of the solution. In
the following Sections we will discuss in detail how to
derive analytic expressions for the limiting relative
velocities and use them efficiently within a numerical
code.

\section{Limiting Relative Velocities: Fundamental Expressions}
\label{lrvs}

	The expression for the relative velocity
(\ref{vrel}) is a relativistic invariant, but its
calculation can be considerably simplified when performed
in an appropriate reference frame. In practice, we will
consider each of the three wave patterns {\it (i)--(iii)}
as being composed of two ``discontinuity fronts'' (two
shocks, two rarefaction waves, or one of each) moving
in opposite directions and separated by a region where a
contact discontinuity is present. In the case of a shock,
in particular, it is useful to use a reference frame
comoving with the shock front, in which the relativistic
expression for the relative velocities ahead of ($a$) and
behind ($b$) the shock takes the form (Taub 1978)
\begin{equation}
\label{rel_shock}
v_{ab}\equiv\frac{v_a-v_b}{1-v_a v_b}
	=\sqrt{\frac{(p_b-p_a)(e_b-e_a)}{(e_a+p_b)(e_b+p_a)}}
	\ .
\end{equation}

	In the case of a rarefaction wave, on the other
hand, it is more convenient to use the Eulerian frame in
which the initial states are measured and which has one
of the axes aligned with the direction of propagation of
the wave front. In such a frame, the flow velocity at the
back of a rarefaction wave (i.e. behind the tail of the
rarefaction wave) can be expressed as a function of
pressure at the back of the wave as
\begin{equation}
\label{raref}
v_b=\frac{(1+v_a)A_{\pm}(p_b)-(1-v_a)}
	{(1+v_a)A_{\pm}(p_b)+(1-v_a)} \ .
\end{equation}
The quantity $A_{\pm}(p)$ in (\ref{raref}) is defined as
(Mart\'{\i} \& M\"uller 1994)
\begin{equation}
\label{apm}
A_{\pm}(p)\equiv\left\{\left[\frac{(\gamma-1)^{1/2}-c_s(p)}
	{(\gamma-1)^{1/2}+c_s(p)}\right]
	\left[\frac{(\gamma-1)^{1/2}+c_s(p_a)}
	{(\gamma-1)^{1/2}-c_s(p_a)}
	\right]\right\}^{\pm2/(\gamma-1)^{1/2}}\ ,
\end{equation}
with the $\pm$ signs corresponding to rarefaction waves
propagating to the left (${\cal R}_\leftarrow$) and to
the right (${\cal R}_\rightarrow$) of the contact
discontinuity, respectively. The quantity $c_s(p)$ in
(\ref{apm}) is the local sound speed which, for a
polytropic equation of state, can be written as
\begin{equation}
\label{sound_speed}
c_s=\sqrt{\frac{\gamma(\gamma-1)p}
	{(\gamma-1)\rho+\gamma p}}\ .
\end{equation}

	We can now use expression (\ref{raref}) to write
an invariant expression for the relative velocity across
a rarefaction wave (i.e. the relative velocity of the
fluid ahead of the rarefaction wave and
behind the tail of the rarefaction wave) as
\begin{equation}
\label{rel_raref}
v_{ab}= \frac{1-A_{\pm}(p_b)}{1+A_{\pm}(p_b)} \ .
\end{equation}

	Expressions (\ref{rel_shock}) and
(\ref{rel_raref}) are not yet in a useful form since they
cannot be combined to give (\ref{vrel}). However, we can
exploit the fact they are relativistic invariants to
evaluate them in the rest frame of the contact
discontinuity. The latter is, by definition, comoving
with the fluid behind the discontinuity fronts and within
such a frame we can set $v_b=0$ and use equations
(\ref{rel_shock}), (\ref{rel_raref}) to obtain an
explicit expression for the velocity ahead of the wave
front. The expressions for the velocities ahead of both
discontinuity fronts obtained in this way can then be
combined so as to evaluate expression (\ref{vrel}).

	In the following Sections we will apply the
procedure outlined above to derive the relative velocity
between the left and right states for the three different
wave patterns {\it (i)--(iii)}. We will denote with
indices $3$ and $3'$ the quantities evaluated in the left
($L_*$) and right ($R_*$) states behind the discontinuity
fronts, and where $p_3=p_{3'}$, $v_3=v_{3'}$, and $\rho_3
\ne \rho_{3'}$. Note also that in all of the different cases
we will assume that $p_1>p_2$, and take the positive
$x$-direction as being the one from the region $1$ to the
region $2$ (An alternative opposite choice is also
possible.).

\section{$L{\cal S}_\leftarrow L_*{\cal
C}R_*{\cal S}_\rightarrow R$: Two Shock Fronts}
\label{case_i}

\begin{figure}[htb]
\centerline{
\psfig{file=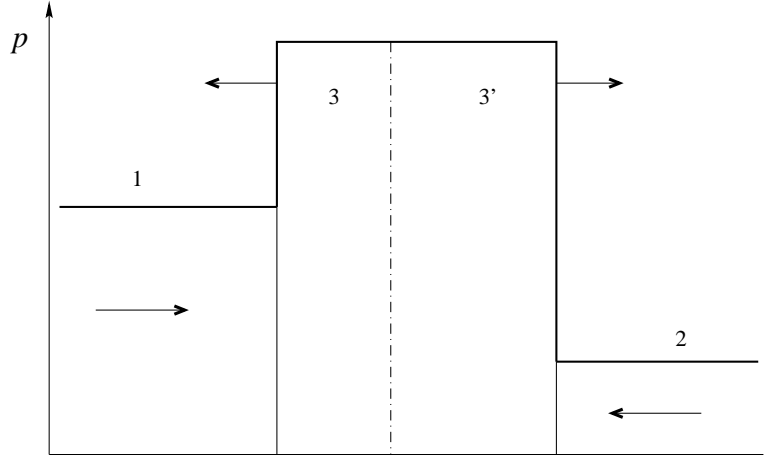,angle=0,width=10.0cm}
        }
\caption{
\label{fig2}
Schematic wave pattern in the pressure for the decay of a
discontinuity generating two shock waves propagating in
opposite directions. The vertical solid lines show the
position of the shock fronts while the dot-dashed
vertical line shows the position of the contact
discontinuity. The different arrows show the gas flow and
the directions of propagation of the different fronts.}
\end{figure}

	We start by considering the wave pattern produced
by two shocks propagating in opposite directions (see
Fig.~\ref{fig2}). This situation is characterized by a
value of the pressure downstream of the shocks which is
larger than the pressures in the unperturbed states,
i.e. $p_3 > p_1 > p_2$. Applying equation
(\ref{rel_shock}) to the shock front moving towards the
left and evaluating it in the reference frame of the
contact discontinuity, we can write the velocity ahead of
the left propagating shock as
\begin{equation}
\label{v1_2s}
v_1=\sqrt{\frac{(p_3-p_1)(e_3-e_1)}
	{(e_1+p_3)(e_3+p_1)}}\ .
\end{equation}

	Similarly, we can apply equation
(\ref{rel_shock}) to the shock front moving towards the
right and evaluate it in the frame comoving with the
contact discontinuity to obtain that the velocity ahead
of the right propagating shock is
\begin{equation}
\label{v2_2s}
v_2=-\sqrt{\frac{(p_3-p_2)(e_{3'}-e_2)}{(e_2+p_3)(e_{3'}+p_2)}}
\end{equation}

	Equations (\ref{v1_2s}) and (\ref{v2_2s}) can now
be used to derive the relativistic expression for the
relative velocity of the flow ahead of the two shocks
$(v_{12})_{_{2S}}$. As proved in Appendix A, the
expression for the relative velocity between the
unperturbed states is a monotonic function of $p_3$ for
all possible wave patterns. In particular, for the
present choice of initial data, this expression is a
monotonically increasing function of $p_3$. As a result,
the value of $(v_{12})_{_{2S}}$ can be used to build a
criterion for the occurrence of two shocks propagating in
opposite directions. In fact, since $p_1$ is the smallest
value that $p_3$ can take, two shocks will form if
\begin{equation}
\label{cond_1}
v_{12} > ({\tilde v}_{12})_{_{2S}}\equiv
	\sqrt{\frac{(p_1-p_2)(\hat{e}-e_2)}
	{(\hat{e}+p_2)(e_2+p_1)}} \ ,
\end{equation}
where 
\begin{equation}
\hat{e}=\hat h \hat \rho - p_1
	= \hat h \frac{\gamma p_1}{(\gamma-1)(\hat h-1)}-p_1 \ ,
\end{equation}
and $\hat h$ is the only positive root of the Taub
adiabat (Taub 1978, Mart\'{\i} \& M\"uller 1994)
\begin{equation}
\label{taub_a}
\left[1+\frac{(\gamma-1)(p_2-p_3)}{\gamma p_3}\right]\hat{h}^2
	-\frac{(\gamma-1)(p_2-p_3)}{\gamma p_3}\hat h 
	+\frac{h_2 (p_2-p_3)}{\rho_2}-h_2^2 = 0 \ .
\end{equation}
when $p_3 \rightarrow p_1$. Simple calculations reported
in Appendix B show that the Newtonian limit of
$({\tilde v}_{12})_{_{2S}}$ corresponds to the
expression derived by Landau and Lifshitz (1987).

\section{$L{\cal S}_\leftarrow L_*{\cal
C}R_*{\cal R}_\rightarrow R$: One Shock and one Rarefaction
Wave}
\label{case_ii}
	
\begin{figure}[htb]
\centerline{
\psfig{file=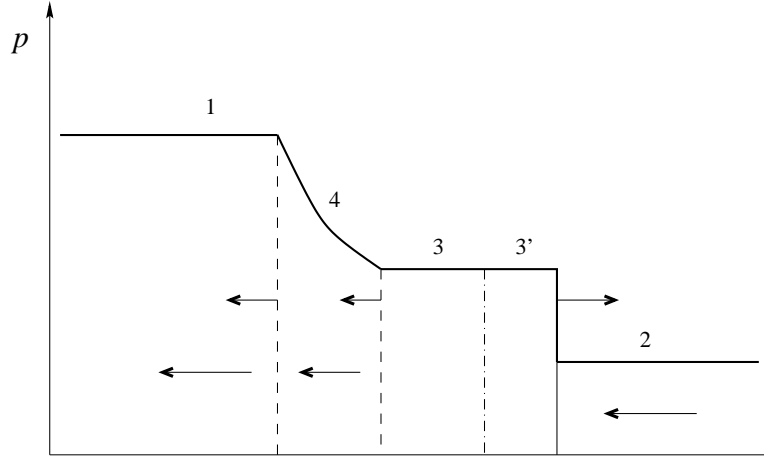,angle=0,width=10.0cm}
        }
\caption{
\label{fig3}
Schematic wave pattern in the pressure for the decay of
the discontinuity into a shock wave propagating towards
the right and a rarefaction wave propagating in the
opposite direction. The vertical lines show the
discontinuities formed (continuous for the shock front;
dashed for the head and the tail of the rarefaction wave;
dot-dashed for the contact discontinuity) while the
arrows show their direction of propagation and that of
the gas flow.}
\end{figure}

	We next consider the wave pattern produced by one
shock front propagating towards the right and one
rarefaction wave propagating in the opposite direction
(see Fig.~\ref{fig3}). This situation is therefore
characterized by $p_1 > p_3 > p_2$. 

	Evaluating expression (\ref{rel_raref}) in the
reference frame comoving with the contact discontinuity,
we can evaluate the flow velocity ahead of the
rarefaction wave to be
\begin{equation}
\label{v1_sr}
v_1=\frac{1-A_{+}(p_3)}{1+A_{+}(p_3)} \ ,
\end{equation}
where
\begin{equation}
\label{a+p3}
A_{+}(p_3) \equiv \left\{\left[\frac{(\gamma-1)^{1/2}-c_s(p_3)}
	{(\gamma-1)^{1/2}+c_s(p_3)}\right]
	\left[\frac{(\gamma-1)^{1/2}+c_s(p_{1})}
	{(\gamma-1)^{1/2}-c_s(p_{1})}
	\right]\right\}^{2/(\gamma-1)^{1/2}}\ .
\end{equation}

	The flow velocity ahead of the shock front can be
derived as in Section \ref{case_i} by evaluating
equation (\ref{rel_shock}) in the reference frame of the
contact discontinuity to get
\begin{equation}
\label{v2_sr}
v_2=-\sqrt{\frac{(p_3-p_2)(e_{3'}-e_2)}
	{(e_{3'}+p_2)(e_2+p_3)}}\ , 
\end{equation}
which, combined with expression (\ref{v1_sr}), can be
used to derive the relativistic expression for the
relative velocity of the fluids ahead of the shock and
ahead of the rarefaction wave $(v_{12})_{_{SR}}$.  As for
$(v_{12})_{_{2S}}$, it can be shown that
$(v_{12})_{_{SR}}$ is a monotonically increasing function
of $p_3$ (see Fig.~\ref{fig1} and Appendix A for an
analytic proof). Exploiting now the knowledge that for
this wave pattern the pressure in the region between the
two waves must satisfy $p_2 < p_3 < p_1$, we can
establish that the criterion on the relative velocity for
having one shock and one rarefaction wave is
\begin{equation}
\label{cond_2}
({\tilde v}_{12})_{_{SR}} = 
	\frac{1-A_{+}(p_3)}{1+A_{+}(p_3)}\bigg\vert_{p_3=p_2} 
	< v_{12}
	\le \sqrt{\frac{(p_1-p_2)(\hat{e}-e_2)}
	{(e_2+p_1)(\hat{e}+p_2)}} =
	({\tilde v}_{12})_{_{2S}}\ .
\end{equation}
Note that the upper limit of (\ref{cond_2}) coincides
with $({\tilde v}_{12})_{_{2S}}$, which is the lower
limit for the occurrence of 2 shock waves (\ref{cond_1})
and whose Newtonian limit coincides with the equivalent
one found by Landau and Lifshitz (1987) (see Appendix
B). Note also that in the limit, $p_3\rightarrow p_2$,
regions 1 and 2 are connected by a single rarefaction
wave. In this case the sound speed can be computed using
$p_3=p_2$ but with $\rho_3=\rho_1(p_2/p_1)^{1/\gamma}$.
Finally, note that this is the only wave
pattern in which $v_1$ and $v_2$ have the same sign and
it therefore includes the classical {\it shock-tube
problem}, where $v_1 = v_2 = 0$.

\section{$L{\cal R}_\leftarrow L_*{\cal
C}R_*{\cal R}_\rightarrow R$: Two Rarefaction Waves}
\label{case_iii}

\begin{figure}[htb]
\centerline{
\psfig{file=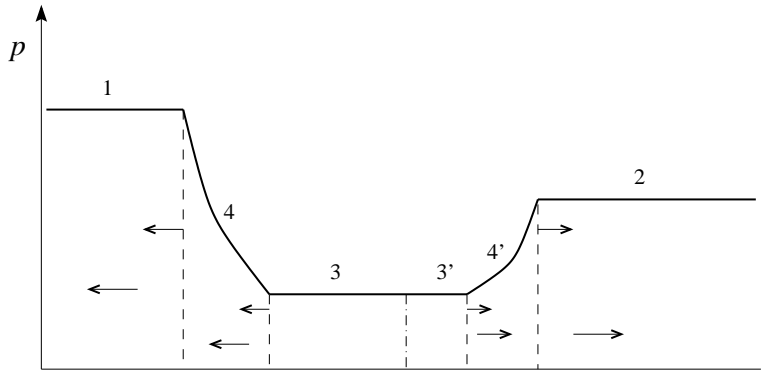,angle=0,width=10.0cm}
        }
\caption{
\label{fig4}
Schematic wave pattern in the pressure for the decay of
the discontinuity into two rarefaction waves propagating in
opposite directions. The vertical lines show the
discontinuities formed (dashed for the head and the tail
of the rarefaction waves; dot-dashed for the contact
discontinuity) while the arrows show their direction of
propagation and that of the gas flow. Note that the
region downstream of the two rarefaction waves has a
density $\rho_3 > 0$.}
\end{figure}

	We now consider the wave pattern produced by two
rarefaction waves propagating in opposite directions (see
Fig.~\ref{fig4}). This situation is characterized by $p_1
> p_2 > p_3$ and when the waves are sufficiently strong
it might lead to a vacuum region ($\rho_3=0$) behind
the rarefaction waves.

	Following again the same procedure discussed in
the previous Section, we can determine the values of the
fluid velocities ahead of the two rarefaction waves as,
respectively,
\begin{eqnarray}
\label{v1_2r}
v_1&=&-\frac{A_{+}(p_3)-1}{A_{+}(p_3)+1}\ ,
\\\nonumber\\
\label{v2_2r}
v_2&=&\frac{1-A_{-}(p_{3'})}{1+A_{-}(p_{3'})}\ ,
\end{eqnarray} 
where 
\begin{equation}
\label{a-p3}
A_{-}(p_{3'}) \equiv \left\{\left[\frac{(\gamma-1)^{1/2}-
	c_s(p_{3'})}{(\gamma-1)^{1/2}+c_s(p_{3'})}\right]
	\left[\frac{(\gamma-1)^{1/2}+c_s(p_{2})}
	{(\gamma-1)^{1/2}-c_s(p_{2})}
	\right]\right\}^{-2/(\gamma-1)^{1/2}} 
	\ . 
\end{equation}
We have indicated with $c_s(p_{3'})$ the sound speed in
the region $3'$, which differs from the one in region $3$
because of the jump in the densities $\rho_3$ and
$\rho_{3'}$. The relative velocity built using
(\ref{v1_2r}) and (\ref{v2_2r}) is then
\begin{equation}
\label{v12_2r}
(v_{12})_{_{2R}} = - \frac{A_{+}(p_3) - A_{-}(p_{3'})}
	{A_{+}(p_3) + A_{-}(p_{3'})} \ .
\end{equation}

	As for the relative velocities of the previous
wave patterns, it can be shown that $(v_{12})_{_{2R}}$,
is a monotonically increasing function of $p_3$ (see
Fig.~\ref{fig1} and Appendix A for an analytic proof) so
that the criterion for the occurrence of two rarefaction
waves can be expressed as
\begin{equation}
\label{cond_3}
	(v_{12})_{_{2R}}\big\vert_{p_3=0}
	< v_{12}
	\le (v_{12})_{_{2R}}\big\vert_{p_3=p_2} 
	= ({\tilde v}_{12})_{_{SR}} \ ,
\end{equation}
where $A_-(p_3=p_2)=1$ and therefore the upper limit for
(\ref{cond_3}) coincides with the lower limit for
(\ref{cond_2}). 

	The condition (\ref{cond_3}) can also be
expressed in a more useful form as
\begin{equation}
- \frac{S_1-S_2}{S_1+S_2} 
	< v_{12}
	\le - \frac{1 - A_{+}(p_2)}{1 + A_{+}(p_2)} \ ,
\end{equation}
where, the constants $S_1$ and $S_2$ are shorthand for 
\begin{eqnarray}
\label{k1}
S_1&\equiv&\left[\frac{(\gamma-1)^{1/2}+c_s(p_1)}
	{(\gamma-1)^{1/2}-c_s(p_1)}\right]^{2/(\gamma-1)^{1/2}}
\\\nonumber\\
\label{k2}
S_2&\equiv&\left[\frac{(\gamma-1)^{1/2}+c_s(p_2)}{(\gamma-1)^{1/2}
	-c_s(p_2)}\right]^{-2/(\gamma-1)^{1/2}} \ .
\end{eqnarray}

	An important property of equation (\ref{v12_2r})
is that it can be inverted analytically. This involves
rewriting it in terms of a quartic equation in the
unknown sound speed in the region $L_*$ (see Appendix C
for the explicit form of the equation). Once the relevant
real root of this equation has been calculated analytically,
the value of the pressure $p_*$ can be found through a
simple algebraic expression. In this case, therefore, the
solution of the exact relativistic Riemann problem can be
found in an analytic closed form.

	We conclude the analysis of the different wave
patterns with a comment on the case of two rarefaction
waves propagating in opposite directions and leaving
behind them a region with zero density and pressure. This
situation occurs when the fluids in the regions 1 and 2
are moving sufficiently fast in opposite directions.
This is the case whenever the relative velocity between the
two states ahead of the rarefaction waves is less than or
equal to the lower limit for $(v_{12})_{_{2R}}$, i.e.
\begin{equation}
v_{12} \le ({\tilde v}_{12})_{_{2R}} 
	\equiv -\frac{S_1-S_2}{S_1+S_2} \ .
\end{equation}
Also in this case, taking the Newtonian limit of
$({\tilde v}_{12})_{_{2R}}$ we obtain the corresponding
expression derived by Landau and Lifshitz (1987) (see
Appendix B). Finally, note that when a vacuum is
produced, $v_{12}$ is no longer dependent on $p_3$ and
this branch of the curve cannot be plotted in
Fig.\ref{fig1}.

\section{Numerical Implementation}
\label{ni}

	The core of most exact Riemann solvers, in both
Newtonian and relativistic hydrodynamics, is based on the
numerical computation of the pressure in the regions
$L_*$ and $R_*$ that form behind the waves. The key
property exploited when performing the numerical
calculation is that the velocity in such regions can be
expressed as a monotonic function of the pressure
i.e. $v_{L_*} = v_{L_*}(p_{L_*})$, and
$v_{R_*}=v_{R_*}(p_{R_*})$. Since there is no jump across
a contact discontinuity in either the velocity or in the
pressure, the numerical solution of the Riemann problem
consists then of finding the root $p_*=p_{L_*}=p_{R_*}$
of the nonlinear equation
\begin{equation}
\label{marti}
v_{L_*}(p_*)-v_{R_*}(p_*) = 0 \ ,
\end{equation}
where $v_{L_*}, v_{R_*}$ have different functional forms
according to the different wave patterns produced. This
method has two obvious disadvantages: {\it (1)} it cannot
determine, using the initial conditions, the wave pattern
produced and thus which of the functional forms to use for
$v_{L_*}, v_{R_*}$; {\it (2)} it cannot provide a
straightforward bracketing interval for the root. In
practice, however, these difficulties are effectively
balanced by efficient algorithms based on a sequence of
trial and error attempts that rapidly bracket the root
and determine the correct equation to solve (see, for
instance, Mart\'{\i}, \& M\"uller 1999 and the algorithm
presented therein).

	The exact Riemann solver which we propose here
differs from the one discussed above mostly because it
avoids the disadvantages {\it (1)} and {\it (2)}. In
fact, as discussed in Section~\ref{intro}, by comparing
the relative velocity between the initial left and right
states $(v_{12})_0$ with the relevant limiting values
constructed from the initial conditions $({\tilde
v}_{12})_{_{2S}}$, $({\tilde v}_{12})_{_{SR}}$, $({\tilde
v}_{12})_{_{2R}}$, we can determine both the wave pattern
which will be produced and the correct bracketing range
in the pressure. Once this information has been obtained,
the Riemann problem can be solved either through the
solution of equation (\ref{marti}) or, equivalently, by
looking for the value of the pressure $p_*$ which would
produce a relative velocity $(v_{12})_0$. This latter
approach, which we will be discussing in the following,
involves then the solution of the nonlinear equation
\begin{equation}
\label{our}
v_{12}(p_*) - (v_{12})_0 = 0 \ ,
\end{equation}
where $v_{12}(p_*)$ is given by the expressions for
$(v_{12})_{_{2S}}$, or $(v_{12})_{_{SR}}$, or
$(v_{12})_{_{2R}}$ derived in
Sections~\ref{case_i}--\ref{case_iii}. Furthermore, in
the case of two rarefaction waves, equation (\ref{our})
can also be solved analytically.

	Besides providing direct information about the
wave pattern produced, about the correct equation to
solve and the relevant bracketing interval, our
approach is also very simple to implement numerically. In
practice, the basic steps for the solution of the Riemann
problem can be summarised as follows:

\begin{enumerate}

\item Evaluate from the initial conditions the three limiting
relative velocities $({\tilde v}_{12})_{_{2S}}$,
$({\tilde v}_{12})_{_{SR}}$, $({\tilde v}_{12})_{_{2R}}$.

\item Determine the wave pattern and the functional form
of $v_{12}(p_*)$ by comparing $(v_{12})_0$ with the
limiting values calculated in {\it (a)} and according to
the scheme below.

\begin{center}
\begin{tabular}{rllll}
& & & & \\
$(v_{12})_0$ & $>$ & $  ({\tilde v}_{12})_{_{2S}}$: & 
	$\ \ {\cal S}_\leftarrow {\cal C}
	{\cal S}_\rightarrow,\ \ $& 
	$v_{12}(p_*)=({v}_{12})_{_{2S}}$\\
$({\tilde v}_{12})_{_{SR}}< (v_{12})_0$ & $\leq$ &$ ({\tilde
	v}_{12})_{_{2S}}$: &  
	$\ \ L{\cal R}_\leftarrow L_*{\cal
	C}R_*{\cal S}_\rightarrow R, \ \ $& 
	$v_{12}(p_*)=({v}_{12})_{_{SR}}$\\
$({\tilde v}_{12})_{_{2R}}< (v_{12})_0$ &$\leq$ &$({\tilde
	v}_{12})_{_{SR}}$: &  
	$\ \ L{\cal R}_\leftarrow L_*{\cal C}R_*
	{\cal R}_\rightarrow R,\ \ $&  
	$v_{12}(p_*)=({v}_{12})_{_{2R}}$\\
$(v_{12})_0$ &$\leq$ & $({\tilde v}_{12})_{_{2R}}$: &  
	$\ \ L{\cal R}_\leftarrow L_*{\cal C}R_*
	{\cal R}_\rightarrow R$ with vacuum,\ \ &
	$\qquad \quad \ - $\\
& & & & \\
\end{tabular}
\end{center}

\item According to the wave pattern found, determine the
extremes $p_{max}$ and $p_{min}$ of the pressure interval
bracketing $p_*$. Within our conventions this is
equivalent to setting\footnote{Note that in practice, the
upper limit for the pressure in the case of two shocks is
found by starting from a reasonable value above $p_{min}$,
which is incremented until the solution is effectively
bracketed.}

\begin{center}
\vskip 0.5truecm
\begin{tabular}{|l|c|c|c|}
	\cline{1-4} & & & \\
&$\ \ {\cal S}_\leftarrow {\cal C}
	{\cal S}_\rightarrow\ \ $	&$\ \ L{\cal R}_\leftarrow L_*{\cal
	C}R_*{\cal S}_\rightarrow R \ \ $ 
	&$\ \ L{\cal R}_\leftarrow L_*{\cal C}R_*
	{\cal R}_\rightarrow R\ \ $
\\ & & & \\ \cline{1-4} & & & \\
 $p_{min}$		& max$(p_1,p_2)$	&min$(p_1,p_2)$
			& 0
\\ & & & \\ \cline{1-4} & & & \\ 
$p_{max}$		& $\infty$		&max$(p_1,p_2)$
			&min$(p_1,p_2)$
\\ & & & \\ \cline{1-4} 
\end{tabular}
\vskip 0.5truecm
\end{center}

\item Solve equation (\ref{our}) and determine $p_*$.

\item Complete the solution of the Riemann problem by
computing the remaining variables of the intermediate
states $L_*$ and $R_*$.

\end{enumerate}

	We have implemented our algorithm for an exact
Riemann solver and have tested it for a range of Riemann
problems. We have also compared the performance of our
algorithm with the ``standard'' approach presented by
Mart\'{\i} \& M\"uller 1999 and have found a systematic
reduction in the computational costs for the same level
of accuracy in the solution. The quantitative efficiency
improvement depends on the type of problem under
consideration. In the case of a generic hydrodynamical
problem (in which very simple Riemann problems are
solved), our approach brackets the solution very closely
and this produces substantial computational improvements
of up to $30\%$. However, for the specific cases
discussed in this paper, where very strong shocks have
been considered, the speed-up is smaller and of the order
of $10\%$. It is worth noting that such an improvement
could reduce appreciably the computational costs in
three-dimensional relativistic hydrodynamics codes, where
this operation is repeated a very large number of times.

\section{Conclusion}
\label{conclusion}

	We have presented a new procedure for the
numerical solution of the exact Riemann problem in
relativistic hydrodynamics. In this approach special
attention is paid to the relativistic invariant
expression for the relative velocity $v_{12}$ between the
unperturbed left and right states. This has been shown to
be a monotonic function of the pressure $p_*$ in the
region formed between the wave fronts. The determination
of this pressure is the basic step in the solution of the
exact Riemann problem.

	The use of the relative velocity has a number of
advantages over alternative exact Riemann solvers
discussed in the literature. In particular, it extracts
the information implicitly contained in the data for two
initial states to deduce the wave pattern that will be
produced by the decay of the discontinuity between these
two states. This, in turn, allows an ``a priori''
determination to be made of the interval in pressure
bracketing $p_*$ and the correct functional form for the
nonlinear equation whose root will solve the exact
Riemann problem. All of these advantages translate, in
practice, into a simpler algorithm to implement and an
improved efficiency in the numerical solution of the
Riemann problem.  Furthermore, in the case of two
rarefaction waves propagating in opposite directions (and
in strict analogy with what happens in Newtonian
hydrodynamics), the use of the relative velocity allows
for the solution of the Riemann problem in an analytic
closed form.

	Because of all of the advantages discussed above,
its intrinsic simplicity of implementation and the
numerical efficiency gain it produces, this new exact
Riemann solver should be considered as an interesting 
alternative to the traditional exact Riemann solver
presently discussed in the literature. Investigations
about the extension of this approach to multidimensions
are currently in progress.
 
\acknowledgements 
It is a pleasure to thank J. A. Font, C. Gheller, and
J. C. Miller for useful discussions. Financial support
for this research has been provided by the Italian
Ministero dell'Universit\`a e della Ricerca Scientifica
and by the EU Programme 'Improving the Human Research
Potential and the Socio-Economic Knowledge Base'
(Research Training Network Contract HPRN-CT-2000-00137)."

\section*{Appendix A: Monotonicity of the relative 
	velocity as function of $p_*$}

	We here prove that $v_{12}$ is a monotonic
function of $p_*$ for all of the possible wave
patterns. In particular, because of our choice of
referring the initial left state as to ``1'' and the
initial right state as to ``2'', we will here show that
$v_{12}$ is a monotonically {\it increasing} function of
$p_*$.

	Denoting by $a'$ the first derivative of the
quantity $a$ with respect to $p_*=p_3=p_{3'}$, it is
straightforward to obtain that the first derivative of
(\ref{vrel}) is
\begin{equation}
\label{vrel'}
v'_{12}= \frac{v'_1(1-v_2^2) - v'_2(1-v_1^2)}{(1-v_1 v_2)^2} \ .
\end{equation}	
Since $v_1\leq 1,\;v_2\leq 1$, the two terms in the
parentheses of (\ref{vrel'}) are always positive, and the
proof that $v_{12}$ is monotonically increasing will
follow if it can be shown that $v'_1$ and $-v'_2$ are
both positive. We will do so for each of the three
possible wave patterns.
\subsection*{Two Shock Fronts}

	Taking the derivative of the fluid velocity ahead of the
left propagating shock front (measured from the contact discontinuity) 
[cf. equation (\ref{v1_2s})] we obtain
\begin{equation}
\label{v_1dot}
v'_1=\frac{(e_1+p_1)\left[(e_3-e_1)(e_3+p_1)+
	(p_3-p_1)(e_1+p_3)e'_3\right]}{2 v_1(e_1+p_3)^2(e_3+p_1)^2}\ .
\end{equation}
Since the energy density is an increasing function of
pressure, $e'_3 > 0 $; furthermore, $p_3\geq p_1$ and
$e_3\geq e_1$ for the wave pattern considered and $v'_1 >
0$ as a result. The equivalent expression for the
derivative of the fluid velocity ahead of the right propagating
shock front (measured from the
contact discontinuity) [cf. eq (\ref{v2_2s})] can be obtained by
replacing in equation (\ref{v_1dot}) the indices $1$ and
$3$ by $2$ and $3'$ respectively, i.e.
\begin{equation}
\label{v_2dot}
v'_2=\frac{(e_2+p_2)\left[(e_{3'}-e_2)(e_{3'}+p_2)+
	(p_{3'}-p_2)(e_2+p_{3'})e'_{3'}\right]}
	{2v_2(e_2+p_{3'})^2(e_{3'}+p_2)^2} \ .
\end{equation}
Since now for the wave pattern considered: $v_2 < 0 $,
$p_{3'}\geq p_2$ and $e_{3'}\geq e_2$, we are led to
conclude that $-v'_2 >0$, thus making the overall
$v'_{12}$ positive for any value of $p_3$.

\subsection*{One Shock and one Rarefaction Wave}

	In this case we only need to show that $v'_1 > 0$
since for the velocity ahead of the right propagating
shock front we can use the results derived in
(\ref{v_2dot}). Taking the derivative of expression
(\ref{v1_sr}) then yields
\begin{equation}
\label{v2_dot}
v'_1=-\frac{2 A'_+(p_3)}{[1+A_+(p_3)]^2} \ ,
\end{equation}
where
\begin{eqnarray}
A'_+(p_3) &=& -4 \bigg\vert\frac{(\gamma-1)^{1/2}+c_s(p_{1})}
	{(\gamma-1)^{1/2}-c_s(p_{1})}\bigg\vert 
	\frac{A_+(p_3)^{(2-\sqrt{\gamma-1})/2}}
	{[(\gamma-1)^{1/2} + c_s(p_3)]^2}{c'_s(p_3)}
	\nonumber \\ \nonumber \\ 
	&\equiv& - C_1 {c'_s(p_3)} \ , 
\end{eqnarray}
where $C_1 >0$. When the sound speed $c_s(p_3)$ is an
increasing function of pressure, as is the case for a
polytropic equation of state [cf. eq (\ref{pol_eos})],
$v'_1>0$ and therefore $v'_{12}>0$.

\subsection*{Two Rarefaction Waves}

	What we need to show in this case is that 
$v'_2<0$ since we can exploit the previous result that
$v'_1>0$ where $v_1$ is the fluid velocity ahead of the
left propagating rarefaction wave as measured from the
contact discontinuity. In this case, taking the
derivative of expression (\ref{v2_2r}) one obtains
\begin{equation}
\label{v3_dot}
v'_2=-\frac{2 A'_-(p_{3'})}{[1+A_-(p_{3'})]^2} \ ,
\end{equation}
where now
\begin{eqnarray}
A'_-(p_{3'}) &=& 4 \bigg\vert\frac{(\gamma-1)^{1/2}+c_s(p_{2})}
	{(\gamma-1)^{1/2}-c_s(p_{2})}
	\bigg\vert \frac{A_-(p_{3'})^{(2+
	\sqrt{\gamma-1})/2}}{[(\gamma-1)^{1/2} + c_s(p_{3'})]^2} 
	c'_s(p_{3'})	\nonumber \\ \nonumber \\ 
	&\equiv& C_2 c'_s(p_{3'})
\end{eqnarray}
with $C_2 > 0 $ and therefore $v'_2 < 0$.

	We have therefore shown that for all of the wave
patterns considered $v'_1 > 0$ and $-v'_2 > 0$, thus
proving that $v_{12}$ is always a monotonically
increasing function of $p_*$.

\section*{Appendix B: Newtonian limits of 
	$({\tilde v}_{12})_{_{2S}}$, $({\tilde
	v}_{12})_{_{SR}}$, $({\tilde v}_{12})_{_{2R}}$}

	We here show that the three limiting values of
$({\tilde v}_{12})_{_{2S}}$, $({\tilde v}_{12})_{_{SR}}$
and $({\tilde v}_{12})_{_{2R}}$ reduce to their Newtonian
counterparts in the limit of $v,c_s \rightarrow 0$, and
$h\rightarrow 1$. In particular, we will restrict
ourselves to considering the case of a polytropic
equation of state (\ref{pol_eos}).

	We start by considering the Newtonian limit of
$({\tilde v}_{12})_{_{2S}}$ which is obtained when $p/e
\ll 1$ and $e\rightarrow 1/V$, with $V=1/\rho$ being the
specific volume. In this case, then
\begin{eqnarray}
({\tilde v}_{12})_{_{2S}}\bigg\vert_{\rm Newt} &=&
	\sqrt{\frac{(p_1-p_2)(\hat{e}-e_2)}
	{\hat{e}e_2}}\nonumber \\ 
	&=& \sqrt{(p_1-p_2)(1/e_2-1/\hat{e})} \ .
\end{eqnarray}
which coincides with the corresponding expression
derived by Landau and Lifshitz (1987) but with inverted
indices.

	We next consider the Newtonian limit of $({\tilde
v}_{12})_{_{SR}}$ which is obtained when both
$c_s(p_3)\ll 1$ and $c_s(p_1)\ll 1$. In this case,
\begin{eqnarray}
A_+(p_3) &\simeq&
	\left(1-\frac{c_s(p_3)}
	{\sqrt{\gamma-1}}\right)^{4/\sqrt{\gamma-1}}
	\left(1+\frac{c_s(p_1)}
	{\sqrt{\gamma-1}}\right)^{4/\sqrt{\gamma-1}}
\nonumber\\
	&\simeq&\left(1-\frac{4c_s(p_3)}{\gamma-1}\right)
        \left(1+\frac{4c_s(p_1)}{\gamma-1}\right)
\nonumber\\
	&\simeq& 1 -\frac{4}{\gamma-1}(c_s(p_3)-c_s(p_1)) \ .
\end{eqnarray}
so that the Newtonian limit of $({\tilde
v}_{12})_{_{SR}}$ is given by
\begin{eqnarray}
({\tilde v}_{12})_{_{SR}}\bigg\vert_{\rm Newt} &=&
	\frac{1 - A_{+}(p_3)}{2}\bigg\vert_{p_3=p_2}\nonumber \\ 
	&\simeq&-\frac{2}{\gamma-1}c_s(p_1)
	\left[1-
	\frac{c_s(p_3)}{c_s(p_1)}\right]\bigg\vert_{p_3=p_2}
	\ .
\end{eqnarray}
Bearing in mind that
\begin{equation}
\frac{c_s(p_3)}{c_s(p_1)}\bigg\vert_{p_3=p_2}=
	\left(\frac{p_2}{p_1}\right)^{(\gamma-1)/2\gamma}
	\ ,
\end{equation}
we obtain
\begin{equation}
({\tilde v}_{12})_{_{SR}}\bigg\vert_{\rm Newt} =
	-\frac{2}{\gamma-1}c_s(p_1)
	\left[1-\left(\frac{p_2}{p_1}
	\right)^{(\gamma-1)/2\gamma}\right] \ ,
\end{equation}
which again coincides with the corresponding expression
derived by Landau and Lifshitz (1987) but with inverted
indices.

	Finally, we consider the Newtonian limit of
$({\tilde v}_{12})_{_{2R}}$ for $c_s(p_1),c_s(p_2)\ll
1$. In this case
\begin{eqnarray}
S_1 &\simeq& 1 + \frac{4 c_s(p_1)}{\gamma-1} \ ,\nonumber \\
S_2 &\simeq& 1 - \frac{4 c_s(p_2)}{\gamma-1}   \ .
\end{eqnarray}
so that the Newtonian limit is given by
\begin{eqnarray}
\label{v12_2r_n}
({\tilde v}_{12})_{_{2R}}\bigg\vert_{\rm Newt} = 
	\frac{S_2-S_1}{S_1+S_2} &=&
	\left[{-\frac{4c_s(p_1)}{\gamma-1}-
	\frac{4c_s(p_2)}{\gamma-1}}\right]
	\left[{2+
	\frac{4}{\gamma-1}[c_s(p_1)-c_s(p_2)]}\right]^{-1}\nonumber\\ 
&=&\left[-\frac{2c_s(p_1)}{\gamma-1}-
	\frac{2c_s(p_2)}{\gamma-1}\right]\left[
	1-\frac{2}{\gamma-1}[c_s(p_1)-c_s(p_2)]\right]\nonumber\\ 
&=&-\frac{2c_s(p_1)}{\gamma-1}-\frac{2c_s(p_2)}{\gamma-1}
	\ .
\end{eqnarray}
Once more, expression (\ref{v12_2r_n}) coincides with the
corresponding expression derived by Landau and Lifshitz
(1987) but with inverted indices.

\section*{Appendix C: A closed form solution in the case of Two
Rarefaction Waves}

	As discussed in Sections~\ref{case_iii} and
\ref{ni}, when $({\tilde v}_{12})_{_{2R}} < (v_{12})_0 <
({\tilde v}_{12})_{_{SR}}$, the initial conditions give
rise to two rarefaction waves and it is possible to
derive a closed form solution for the unknown pressure
$p_*$. In this way we can, at least in principle, avoid
any numerical root finding procedure and determine the
solution exactly. In this Appendix we first derive this
analytic solution in the context of relativistic hydrodynamics and then
calculate its Newtonian limit. We will
restrict ourselves to considering the particular case of a
polytropic equation of state (\ref{pol_eos}).

	Using expression (\ref{sound_speed}) we can write
the pressures $p_3$ and $p_{3'}$ as functions of the
sound speeds $c_s(p_3)$ and $c_s(p_{3'})$ which, for
convenience, we will hereafter refer to as $x$ and $x'$
respectively
\begin{eqnarray}
\label{cs1}
p_3 &=& k_1^{-1/(\gamma-1)}\left[\frac{x^2
	(\gamma-1)}{\gamma(\gamma-1)-\gamma
	x^2}\right]^{\gamma/(\gamma-1)}  \ , \\ 
\label{cs2}
p_{3'} &=& k_2^{-1/(\gamma-1)}\left[\frac{(x')^2
	(\gamma-1)}{\gamma(\gamma-1)-\gamma(x')^2}
	\right]^{\gamma/(\gamma-1)}  \ ,
\end{eqnarray}
where $k_1=p_1/\rho_1^{\gamma}$ and $k_2 =
p_2/\rho_2^{\gamma}$ are the two constants entering the
polytropic equation. Since $p_3=p_{3'}=p_*$, we can 
obtain the following relation between $x'$ and $x$:
\begin{equation}
\label{sounds_1}
(x')^2\equiv \frac{\gamma(\gamma-1)x^2}{\gamma
	x^2(1-\alpha)+\alpha\gamma(\gamma-1)} \ ,
\end{equation}
where $\alpha\equiv(k_1/k_2)^{1/\gamma}$. The expression
for the relative velocity (\ref{v12_2r}) can also be
written as 
\begin{equation}
\label{appp}
\frac{A_{+}(p_3)}{A_{-}(p_{3'})}=
	\frac{1-(v_{12})_0}{1+(v_{12})_0}
	\ .
\end{equation}
and we then use expressions (\ref{a+p3}) and (\ref{a-p3}) to
expand the left hand side of (\ref{appp}). After some
algebra we are then left with
\begin{equation}
\label{gasp1}
\left(\frac{\Gamma - x}{\Gamma + x}\right)\left(\frac{\Gamma -x'}
	{\Gamma + x'}\right) =
	\left[\frac{\Gamma - c_s(p_2)}
	{\Gamma + c_s(p_2)}\right]\left[\frac{\Gamma -
	c_s(p_{1})}{\Gamma + c_s(p_{1})}\right]	
	\left[\frac{1-(v_{12})_0}{1+
	(v_{12})_0}\right]^{\Gamma/2} \ ,
\end{equation}
where $\Gamma^2 \equiv \gamma - 1$ and the right hand
side of (\ref{gasp1}) is a constant which we rename as 
\begin{equation}
\Pi\equiv\left[\frac{\Gamma - c_s(p_2)}{\Gamma +
	c_s(p_2)}\right]\left[\frac{\Gamma - c_s(p_{1})}{\Gamma +
	c_s(p_{1})}\right]
	\left[\frac{1-(v_{12})_0}
	{1+(v_{12})_0}\right]^{\Gamma/2} \ .
\end{equation}
Introducing now the auxiliary quantity 
\begin{equation}
\beta\equiv\frac{1+\Pi}{1-\Pi} \ ,
\end{equation}
expression (\ref{gasp1}) can be written as
\begin{equation}
\label{sounds_2}
(x')^2=\left[\frac{\Gamma(x\beta - \Gamma)}
	{x - \Gamma\beta}\right]^2 \ .
\end{equation}
Comparing now (\ref{sounds_1}) and (\ref{sounds_2})
gives a
$4$-th order equation in the unknown sound velocity $x$
\begin{equation}
\label{4_order}
a_0 x^4 + a_1 x^3 + a_2 x^2 + a_3
x + a_4 = 0 \ ,
\end{equation}
where 
\begin{eqnarray}
a_0 &\equiv& 1-\beta^2(1-\alpha) \ ,\\
a_1 &\equiv& -2 \Gamma \alpha\beta \ , \\
a_2 &\equiv& \Gamma^2 (1-\alpha)(\beta^2-1) \ , \\
a_3 &\equiv& 2 \Gamma^3 \alpha \beta \ , \\
a_4 &\equiv& -\alpha \Gamma^4  \ .
\end{eqnarray}
The analytic solution of equation (\ref{4_order}) will
yield at least two real roots, one of which will be the
physically acceptable one: i.e. positive, less than one,
and such that the pressure $p_*$ falls in the relevant
bracketing interval.
	
	In its Newtonian limit, equation (\ref{4_order})
reduces to a second order equation in the unknown sound
velocity
\begin{equation} 
\label{2_order}
\left(\frac{1}{\alpha} - 1\right) x^2 + 
	2 \Sigma x - \Sigma^2 = 0 \ ,
\end{equation}
where 
\begin{equation}
\Sigma \equiv c_s(p_1)+c_s(p_2)+\frac{(\gamma-1)}{2} v_{12} \ ,
\end{equation}
and $v_{12}=v_1-v_2$. The fact that the Newtonian Riemann
problem in the case of two rarefaction waves can be
solved analytically is well known and is at the basis of
the so called ``Two Rarefaction Approximate Riemann
Solver'' (Toro 1997).


\end{document}